\documentclass[letterpaper,twocolumn,american,showpacs,prl,aps,superscriptaddress]{revtex4}
\usepackage[latin1]{inputenc}
\usepackage{bm}
\usepackage{amsmath}
\usepackage{graphicx}
\usepackage{amssymb}
\makeatletter
\usepackage{pifont}
\makeatother

\begin{document}

\title{Optimal entanglement criterion for mixed quantum states}

\author{Heinz-Peter Breuer}

\email{breuer@physik.uni-freiburg.de}

\affiliation{Physikalisches Institut, Universit\"at Freiburg,
             Hermann-Herder-Strasse 3, D-79104 Freiburg, Germany}

\date{\today}

\begin{abstract}
We develop a strong and computationally simple entanglement
criterion. The criterion is based on an elementary positive map
$\Phi$ which operates on state spaces with even dimension $N\geq
4$. It is shown that $\Phi$ detects many entangled states with
positive partial transposition (PPT) and that it leads to a class
of optimal entanglement witnesses. This implies that there are no
other witnesses which can detect more entangled PPT states. The
map $\Phi$ yields a systematic method for the explicit
construction of high-dimensional manifolds of bound entangled
states.
\end{abstract}


\maketitle

Entanglement and quantum inseparability are key features of
quantum mechanics which are connected to the tensor product
structure of the state spaces of composite systems. A mixed state
$\rho$ of a bipartite system, for instance, is defined to be
separable or classically correlated if it can be written as a
convex linear combination of uncorrelated product states, i.~e.,
if it can be represented in the form $\rho=\sum_i p_i \rho_1^i
\otimes \rho_2^i$, where $\{p_i\}$ is a probability distribution
and the $\rho_1^i$, $\rho_2^i$ are density matrices describing
states of the first and the second subsystem, respectively
\cite{WERNER}. States which cannot be written in this way are
called inseparable or entangled. Much effort in quantum
information theory has been devoted to the problems of the
characterization, the classification and the quantification of
mixed state entanglement \cite{ALBER,ECKERT}. Although
considerable progress has been made in recent years (see, e.~g.,
Refs.~\cite{DOHERTY}), we are still far away from a true
understanding of many aspects of these problems.

A problem of central importance in entanglement theory is the
development of computationally efficient criteria which allow us
to decide whether or not a given state is entangled. Peres
\cite{PERES} has developed a very strong entanglement criterion
which is known as criterion of positive partial transposition
(PPT). It states that a necessary condition for a given state
$\rho$ to be separable is that its partial transpose is a positive
operator. Usually, one writes this condition as $(I\otimes T)\rho
\geq 0$, where $T$ denotes the transposition of operators in a
chosen basis and $I$ is the identity map, indicating that the
transposition is carried out only on the second part of the
composite system. The PPT condition represents a necessary and
sufficient separability criterion for certain low-dimensional
systems \cite{HORODECKI96}, but it is only necessary in higher
dimensions. Hence, there are generally entangled PPT states which
belong to the class of bound entangled states \cite{HORODECKI98}.

The transposition $T$ is a distinguished example of a positive but
not completely positive map. This means that $T$ maps all positive
operators on the subsystems to positive operators, while there
exist states $\rho$ of the combined system for which $(I\otimes
T)\rho$ has negative eigenvalues. There are many other maps with
this property. The significance of positive maps in entanglement
theory is provided by a fundamental theorem established in
Ref.~\cite{HORODECKI96}. This theorem states that a necessary and
sufficient condition for a state $\rho$ to be separable is that
$(I\otimes \Lambda)\rho$ is positive for any positive map
$\Lambda$. Hence, a given state is separable if and only if it
remains positive under the application of all positive maps to one
of its local parts.

By virtue of the PPT criterion the development of appropriate
separability criteria reduces to the construction of those
positive maps which are able to detect entangled PPT states. Such
maps are called nondecomposable \cite{WORONOWICZ} because they
cannot be written as a sum of a completely positive map and of the
composition of a completely positive map with the transposition
map. However, this formulation does not lead to a simple
operational entanglement criterion since the general structural
characterization of positive maps is an unsolved mathematical
problem. In particular, the explicit construction of
nondecomposable positive maps turns out to be an extremely
difficult task.

Here, we develop a universal nondecomposable positive map $\Phi$
which operates on the states of any state space with even
dimension $N\geq 4$. The map $\Phi$ is composed of elementary
operations and yields a very strong separability criterion which
is particularly efficient for the identification of entangled PPT
states. We show that $\Phi$ detects many entangled states in
arbitrary dimensions which are neither detected by the PPT
criterion nor by the strong realignment criterion
\cite{CHEN,RUDOLPH}.

It is known that positive maps are in one-to-one correspondence to
certain observables called entanglement witnesses
\cite{LEWENSTEIN00}. The map $\Phi$ constructed here has the
remarkable property that it automatically leads for all $N$ to
entanglement witnesses which have the property of being
{\textit{optimal}}. This means that there are no other witnesses
which are finer, i.~e., which are able to identify more entangled
PPT states. Moreover, we develop a systematic method for the
explicit construction of high-dimensional manifolds of bound
entangled states for arbitrary $N$.

We consider an $N$-state system with Hilbert space ${\mathbb
C}^N$. Without loss of generality, we will regard ${\mathbb C}^N$
as the state space of a particle with spin $j$, where $N=2j+1$.
The corresponding basis states are denoted by $|j,m\rangle$, where
$m=-j,-j+1,\ldots,+j$. Since we assume that $N$ is even the spin
$j$ must be half-integer valued.

An important ingredient of our separability criterion is the
symmetry transformation of the time reversal which is described by
an antiunitary operator $\theta$ \cite{GALINDO}. As for any
antiunitary operator, we can write $\theta=V\theta_0$, where
$\theta_0$ denotes the complex conjugation in the chosen basis
$|j,m\rangle$, and $V$ is a unitary operator. In the spin
representation introduced above the matrix elements of $V$ are
given by $\langle j,m'|V|j,m\rangle=(-1)^{j-m}\delta_{m',-m}$. For
even $N$ this matrix is not only unitary but also skew-symmetric,
i.~e., we have $V^T=-V$, where $T$ denotes the transposition. It
follows that $\theta^2=-I$ which leads to
\begin{equation} \label{ORTHO}
 \langle\varphi|\theta\varphi\rangle = 0.
\end{equation}
This relation expresses a well-known property of the time reversal
transformation $\theta$ which will play a crucial role in the
following: For any state vector $|\varphi\rangle$ of the spin-$j$
particle the time-reversed state $|\theta\varphi\rangle$ is
orthogonal to $|\varphi\rangle$. This is a distinguished feature
of even-dimensional state spaces, because unitary and
skew-symmetric matrices do not exist in state spaces with odd
dimension.

The action of the time reversal transformation on an operator $B$
on ${\mathbb C}^N$ can be expressed by
\begin{equation} \label{DEF-THETA}
 \vartheta B = \theta B^{\dagger} \theta^{-1}
 = V B^T V^{\dagger}.
\end{equation}
This defines a linear map $\vartheta$ which transforms any
operator $B$ into its time reversed operator $\vartheta B$. For
example, if we take the spin operator $\hat{\bm{j}}$ of the
particle Eq.~(\ref{DEF-THETA}) gives the spin flip transformation
$\vartheta \hat{\bm{j}} = -\hat{\bm{j}}$. According to the second
relation in Eq.~(\ref{DEF-THETA}) the map $\vartheta$ is unitarily
equivalent to the transposition map. Hence, the PPT criterion is
equivalent to the condition that the partial time reversal
$\vartheta_2$ is positive:
\[
 \vartheta_2 \rho \equiv (I \otimes \vartheta) \rho \geq 0.
\]

We define a linear map $\Phi$ which acts on operators $B$ on
${\mathbb C}^N$ as follows:
\begin{equation} \label{PHI}
 \Phi B = ({\mathrm{tr}}B)  I - B - \vartheta B.
\end{equation}
It will be demonstrated below that this map is positive but not
completely positive. Hence, it yields the following necessary
condition for separability:
\begin{equation} \label{TRC}
 \Phi_2 \rho \equiv (I \otimes \Phi) \rho \geq 0.
\end{equation}
To motivate definition (\ref{PHI}) we recall that in another
separability criterion, known as reduction criterion
\cite{HORODECKI99,CERF}, one uses the positive map defined by
$\Lambda B = ({\mathrm{tr}}B) I - B$. Comparing this definition
with Eq.~(\ref{PHI}) we see that $\Phi=\Lambda-\vartheta$. Hence,
not only the map $\Lambda$ of the reduction criterion and the time
reversal $\vartheta$ are positive, but also their difference
$\Phi=\Lambda-\vartheta$. The criterion (\ref{TRC}) can therefore
be restated as $\Lambda_2 \rho - \vartheta_2 \rho \geq 0$. If
$\rho$ is a PPT state, i.~e., if $\vartheta_2\rho\geq 0$ we
subtract a positive operator from $\Lambda_2\rho$ when evaluating
condition (\ref{TRC}), which sharpens the condition of the
reduction criterion considerably. For this reason the inequality
(\ref{TRC}) can be expected to yield a very strong separability
criterion which is particularly suited to recognize the
entanglement of PPT states.

It is easy to prove the positivity of the map $\Phi$ defined by
Eq.~(\ref{PHI}). We have to show that for any positive operator
$B$ also the operator $\Phi B$ is positive. This statement is
obviously equivalent to the statement that the operator
$\Phi(|\varphi\rangle\langle\varphi|)$ is positive for any
normalized state vector $|\varphi\rangle$. Using definition
(\ref{PHI}) we find:
\[
 \Phi(|\varphi\rangle\langle\varphi|)
 = I - |\varphi\rangle\langle\varphi|
 - |\theta\varphi\rangle\langle\theta\varphi| \equiv I - \Pi.
\]
Because of Eq.~(\ref{ORTHO}) the operator $\Pi$ introduced here
represents an orthogonal projection operator which projects onto
the subspace spanned by $|\varphi\rangle$ and
$|\theta\varphi\rangle$. It follows that also
$\Phi(|\varphi\rangle\langle\varphi|)$ is a projection operator
and, hence, that it is positive for any normalized state vector
$|\varphi\rangle$. This proves our claim. Note that for $N=2$ the
projection $\Pi$ is identical to the unit operator such that
$\Phi$ is equal to the zero map in this case. For this reason we
restrict ourselves to the cases of even $N\geq 4$.

To show that the map $\Phi$ is not completely positive we consider
the tensor product space ${\mathbb C}^N\otimes{\mathbb C}^N$ of
two spin-$j$ particles. The total spin of the composite system
will be denoted by $J$. According to the triangular inequality $J$
takes on the values $J=0,1,\ldots,2j=N-1$. The projection operator
which projects onto the manifold of states corresponding to a
definite value of $J$ will be denoted by $P_J$. In particular,
$P_0$ represents the one-dimensional projection onto the maximally
entangled singlet state $J=0$. We define a Hermitian operator $W$
by applying the tensor extension of $\Phi$ to the singlet state:
\begin{eqnarray} \label{DEF-W}
 W &\equiv& N(I \otimes \Phi) P_0 \nonumber \\
 &=& -(N-2)P_0 + 2P_2 + 2P_4 + \ldots + 2P_{2j-1}.
\end{eqnarray}
In the second line we have used definition (\ref{PHI}), the fact
that the sum of the $P_J$ is equal to the unit operator, the
relation ${\mathrm{tr}}_2 P_0 = I/N$ (${\mathrm{tr}}_2$ denotes
the partial trace taken over subsystem 2), and the formula
\cite{NtensorN}:
\begin{equation} \label{THETA-P0}
 \vartheta_2P_0 = \frac{1}{N} F = -\frac{1}{N}\sum_{J=0}^{2j} (-1)^JP_J,
\end{equation}
where $F$ denotes the swap operator which is defined by
$F|\varphi_1\rangle \otimes |\varphi_2\rangle = |\varphi_2\rangle
\otimes |\varphi_1\rangle$. We infer from Eq.~(\ref{DEF-W}) that
the operator $W$ has the negative eigenvalue $-(N-2)$
corresponding to the singlet state $J=0$. Therefore, $W$ is not
positive and the map $\Phi$ is not completely positive.

Next we show that the criterion (\ref{TRC}) detects entangled PPT
states for all even $N \geq 4$. To this end, it is again useful to
employ the operator $W$ defined by Eq.~(\ref{DEF-W}). Since $\Phi$
is positive but not completely positive $W$ is an entanglement
witness \cite{HORODECKI96,TERHAL}. We recall that an entanglement
witness is a Hermitian operator which satisfies
${\mathrm{tr}}\{W\sigma\}\geq 0$ for all separable states
$\sigma$, and ${\mathrm{tr}}\{W\rho\}< 0$ for at least one
entangled state $\rho$, in which case we say that $W$ detects
$\rho$. A witness $W$ is called nondecomposable if it can detect
entangled PPT states \cite{LEWENSTEIN00}. We prove that there are
always entangled PPT states $\rho$ which are detected by the
witness defined in Eq.~(\ref{DEF-W}), i.~e., for which
${\mathrm{tr}}\{ W \rho \} < 0$. In other words, our witness $W$
is nondecomposable. This implies that also $\Phi$ is
nondecomposable and that the stronger criterion (\ref{TRC}) always
detects entangled PPT states.

Consider the following one-parameter family of states:
\begin{equation} \label{FAMILY}
 \rho(\lambda) = \lambda P_0 + (1-\lambda) \rho_0,
 \qquad 0 \leq \lambda \leq 1.
\end{equation}
These normalized states are mixtures of the singlet state $P_0$
and of the state
\[
 \rho_0 = \frac{2}{N(N+1)} P_S = \frac{2}{N(N+1)} \sum_{J \; \mathrm{odd}} P_J
\]
which is proportional to the projection $P_S$ onto the subspace of
states which are symmetric under the swap operation. Note that
$\rho_0$ is a separable state which belongs to the class of Werner
states \cite{WERNER}. Since $P_S$ can be written as a sum over the
projections $P_J$ with odd $J$, we immediately get with the help
of Eq.~(\ref{DEF-W}):
\[
 {\mathrm{tr}}\{ W \rho(\lambda)\} = -\lambda(N-2).
\]
Hence, we find that ${\mathrm{tr}}\{ W \rho(\lambda)\} < 0$ for
$\lambda > 0$. We conclude that all states of the family
(\ref{FAMILY}) corresponding to $\lambda > 0$ are entangled, and
that $\rho_0$ is the only separable state of this family. On the
other hand, using the representation $P_S=(I+F)/2$ and
Eq.~(\ref{THETA-P0}) we find
\[
 \vartheta_2 \rho(\lambda) = \frac{1-2\lambda}{N}P_0
 + \frac{1}{N} \sum_{J=1}^{2j}
 \left[ (-1)^{J+1} \lambda + \frac{1-\lambda}{N+1} \right] P_J.
\]
It is not hard to check by means of this equation that the PPT
condition $\vartheta_2\rho(\lambda)\geq 0$ is equivalent to
$\lambda \leq 1/(N+2)$. Hence, all $\rho(\lambda)$ with $0 <
\lambda \leq 1/(N+2)$ are entangled PPT states which are detected
by the witness $W$. This proves that the witness $W$ and the map
$\Phi$ are nondecomposable.

The above argument demonstrates that the inequality
${\mathrm{tr}}\{W\rho\} \geq 0$ represents a necessary and
sufficient separability condition for the family of states
(\ref{FAMILY}). Obviously, this criterion cannot be improved by
introducing other witnesses which leads to the idea that $W$ is an
optimal entanglement witness. To make this idea more precise we
introduce the following notations \cite{LEWENSTEIN00}. We denote
by $D_{W}$ the set of all entangled PPT states of the total state
space which are detected by some given nondecomposable witness
$W$. A witness $W_2$ is said to be finer than a witness $W_1$ if
$D_{W_1}$ is a subset of $D_{W_2}$, i.~e., if all entangled PPT
states which are detected by $W_1$ are also detected by $W_2$. A
given witness is said to be optimal if there is no other witness
which is finer, i.~e., if there is no other witness which is able
to detect more entangled PPT states. It is a remarkable fact that
our witness $W$ is always optimal in this sense.

{\textit{Theorem.}} The operator $W=N(I\otimes\Phi)P_0$ on
${\mathbb C}^N \otimes {\mathbb C}^N$ is a nondecomposable optimal
entanglement witness for all even $N \geq 4$.

{\textit{Proof.}} The proof is based on results of Lewenstein,
Kraus, Horodecki, and Cirac \cite{LEWENSTEIN01}. Following these
authors we define for any given entanglement witness $W$ the set
$\Gamma_W$ as the set of all product vectors
$|\varphi_1,\varphi_2\rangle\equiv|\varphi_1\rangle\otimes|\varphi_2\rangle$
in ${\mathbb C}^N \otimes {\mathbb C}^N$ for which the expectation
value of $W$ is equal to zero, i.~e., for which the relation
\begin{equation} \label{GAMMAW}
 \langle\varphi_1,\varphi_2 |W|\varphi_1,\varphi_2\rangle = 0
\end{equation}
holds. According to Ref.~\cite{LEWENSTEIN01} a given
nondecomposable entanglement witness $W$ is optimal if the
elements of the set $\Gamma_W$ as well as the elements of the set
$\Gamma_{\vartheta_2W}$ span the total state space ${\mathbb C}^N
\otimes {\mathbb C}^N$. In the present case we have
$\vartheta_2W=W$ which follows from the relation $\vartheta \Phi =
\Phi$ [see Eq.~(\ref{PHI})]. Hence, we only have to show that the
elements of $\Gamma_W$ corresponding to our witness $W$ span the
state space of the composite system.

The elements of $\Gamma_W$ can easily be characterized. We take
any normalized product vector
$|\varphi_1\rangle\otimes|\varphi_2\rangle$ and use definitions
(\ref{DEF-W}) and (\ref{PHI}) to evaluate the condition
(\ref{GAMMAW}):
\[
  \langle\varphi_1,\varphi_2 |W|\varphi_1,\varphi_2\rangle
  = 1 - |\langle\varphi_1|\varphi_2\rangle|^2
  - |\langle\varphi_1|\theta\varphi_2\rangle|^2 = 0.
\]
This equation is fulfilled if and only if $|\varphi_1\rangle$ lies
in the subspace spanned by the orthogonal vectors
$|\varphi_2\rangle$ and $|\theta\varphi_2\rangle$. In particular,
all product vectors of the form
\begin{equation} \label{PRODUCT-VECTORS}
 |\varphi\rangle \otimes |\theta\varphi\rangle
 \;\;\; {\mbox{or}} \;\;\;
 |\theta\varphi\rangle \otimes |\varphi\rangle
\end{equation}
belong to $\Gamma_W$, where $|\varphi\rangle \in {\mathbb C}^N$ is
arbitrary.

Consider now an arbitrary product vector
$|\phi_1\rangle\otimes|\phi_2\rangle$, and define
$|\varphi_1\rangle=|\theta\phi_1\rangle+|\phi_2\rangle$ and
$|\varphi_2\rangle=i|\theta\phi_1\rangle+|\phi_2\rangle$. Then one
can easily check the following identity:
\begin{eqnarray*}
 \lefteqn{ |\phi_1\rangle\otimes|\phi_2\rangle = } \\
 && -\frac{1}{2}|\theta\varphi_1\rangle\otimes|\varphi_1\rangle
 -\frac{i}{2}|\theta\varphi_2\rangle\otimes|\varphi_2\rangle \\
 && -\frac{1}{2}(1+i)|\phi_1\rangle\otimes|\theta\phi_1\rangle
    +\frac{1}{2}(1+i)|\theta\phi_2\rangle\otimes|\phi_2\rangle.
\end{eqnarray*}
The right-hand side of this identity is a linear combination of
four product vectors of the form (\ref{PRODUCT-VECTORS}). We
conclude that any product vector
$|\phi_1\rangle\otimes|\phi_2\rangle$ can be represented as a
linear combination of elements of $\Gamma_W$. Since any state
vector in ${\mathbb C}^N\otimes{\mathbb C}^N$ can of course be
written as linear combination of product vectors, this implies
that any state vector can be represented as linear combination of
elements of $\Gamma_W$. In other words, the set $\Gamma_W$ indeed
spans the whole Hilbert space, which proves the theorem.

Due to its optimality the separability criterion (\ref{TRC}) can
be much stronger than other known separability criteria. Let us
illustrate this point by means of the family of states defined by
Eq.~(\ref{FAMILY}). Each separability criterion recognizes the
entanglement of the states $\rho(\lambda)$ with $\lambda^c <
\lambda \leq 1$, where $\lambda^c$ is a certain threshold value
depending on the criterion chosen. The weakest criterion is the
reduction criterion which gives $\lambda^c=1/N$. The same result
is obtained if one uses the majorization criterion \cite{KEMPE} or
the quantum R\'{e}nyi entropy $S_{\infty}$ \cite{HORODECKI99}.
Surprisingly, the realignment criterion, which is known to be able
to recognize many entangled PPT states, is not stronger than the
reduction criterion in the present case, i.~e., we again have
$\lambda^c=1/N$. The PPT criterion is slightly better and yields
$\lambda^c=1/(N+2)$. As shown above, the most efficient criterion
is obtained by means of the map $\Phi$ which leads to the optimal
value $\lambda^c=0$.

A further instructive example is given by the set of rotationally
symmetric states \cite{SU2} on the state space ${\mathbb
C}^4\otimes{\mathbb C}^4$. These are the states of a system which
is composed of two particles with spin $j=3/2$ and which is
invariant under unitary product representations of the group
SU(2). As shown in \cite{NtensorN} the map $\Phi$ detects
{\textit{all}} entangled PPT states in this case, i.~e., the
inequality $\Phi_2 \rho \geq 0$ taken together with the PPT
criterion $\vartheta_2 \rho \geq 0$ represents a necessary
{\textit{and}} sufficient separability condition for all
SU(2)-invariant states.

The map $\Phi$ is not only useful in detecting entangled PPT
states but also provides us with a simple and systematic method of
constructing high-dimensional manifolds of such states for
arbitrary dimensions $N$. We take any entangled PPT state
$\rho_{\mathrm{ppt}}$ which is detected by $W$, e.~g., a PPT state
of the family (\ref{FAMILY}). Then
\begin{equation} \label{PPT-MANIFOLD}
 \rho = \rho_{\mathrm{ppt}}
 + \sum_{\alpha} p_{\alpha}
 |\varphi^{\alpha}_1,\varphi^{\alpha}_2\rangle
 \langle\varphi^{\alpha}_1,\varphi^{\alpha}_2|
\end{equation}
is again an (unnormalized) entangled PPT state, where $p_{\alpha}
\geq 0$ and the sum is extended over an arbitrary collection of
product vectors $|\varphi^{\alpha}_1,\varphi^{\alpha}_2\rangle$
taken from $\Gamma_W$. We have a large freedom in the choice of
these vectors: The only condition is that for each $\alpha$ the
state $|\varphi^{\alpha}_1\rangle$ lies in the subspace which is
spanned by $|\varphi^{\alpha}_2\rangle$ and
$|\theta\varphi^{\alpha}_2\rangle$. For example, identifying the
index $\alpha$ with the quantum number $m$ we can choose
$|\varphi^{\alpha}_2\rangle=|j,m\rangle$ and
$|\varphi^{\alpha}_1\rangle=|j,m\rangle$ or
$|\varphi^{\alpha}_1\rangle=|j,-m\rangle$. Equation
(\ref{PPT-MANIFOLD}) then represents a $2N$-dimensional manifold
of entangled PPT states.

Summarizing, we have constructed a universal nondecomposable
positive map which leads to a powerful separability criterion and
to a class of optimal entanglement witnesses. Our results suggest
many further studies and applications. An important issue, for
example, is the investigation of the properties of entanglement
measures. Recently, Chen, Albeverio, and Fei \cite{ALBEVERIO} have
derived lower bounds for the concurrence \cite{WOOTTERS} and for
the entanglement of formation \cite{BENNETT} by connecting these
entanglement measures with the PPT criterion and the realignment
criterion. It is very likely that this connection can be extended
to the optimal entanglement criterion developed here, which will
yield a considerable improvement of the known analytical bounds
for entanglement measures.

\end{document}